\documentclass[runningheads]{llncs}
\usepackage{hyperref}
\usepackage{graphicx}
\usepackage{amsmath,amssymb}
\usepackage{array, tabularx}
\usepackage{thmtools,thm-restate}
\usepackage{algorithm}
\usepackage[noend]{algpseudocode}
\usepackage{color}
\usepackage[noadjust]{cite}
\usepackage{multirow}
\usepackage{caption}
\usepackage{subcaption}
\usepackage{todonotes}
\usepackage{calc}
\usepackage{complexity}
\usepackage{lipsum}
\usepackage{paralist}

\renewcommand{\orcidID}[1]{\href{https://orcid.org/#1}{\includegraphics[scale=.03]{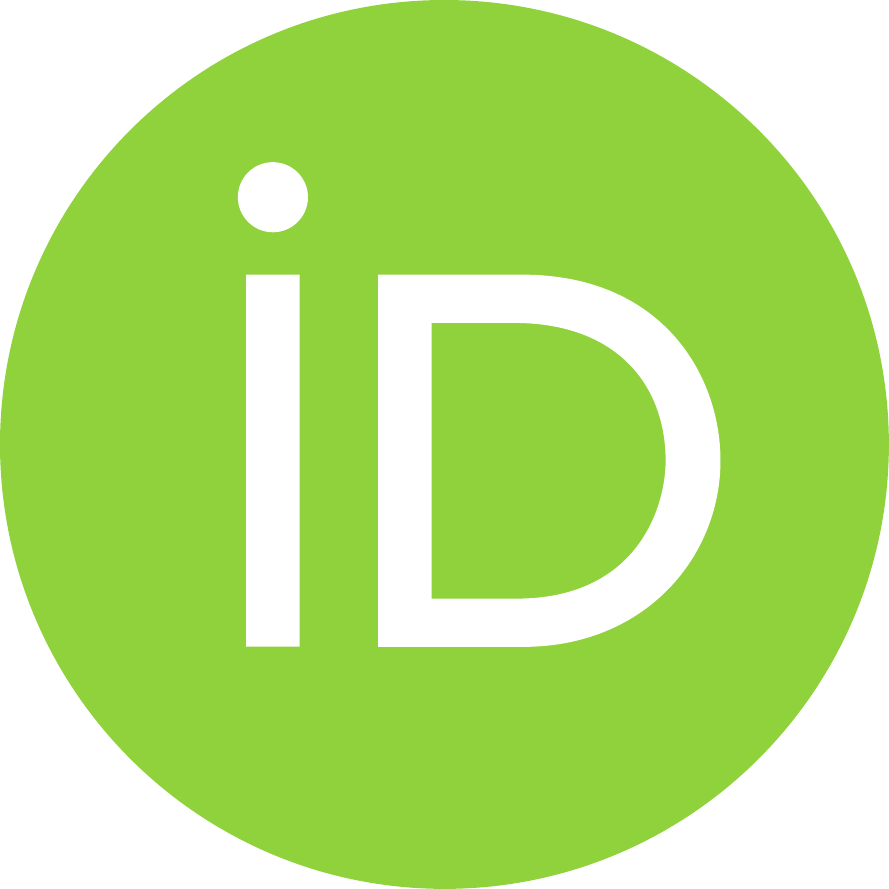}}} 

\hypersetup{
hidelinks,
colorlinks=true,
citecolor=[rgb]{0.121 0.47 0.705},
linkcolor=[rgb]{0.7 0 0.4},
urlcolor=[rgb]{0.121 0.47 0.705}
}

\newcommand{\orientation}[1]{\textsc{orientation}(#1)}
\newcommand{\lab}[1]{\textsc{label}(#1)}
\newcommand{\coordinates}[1]{\textsc{coordinates}(#1)}
\newcommand{\incoordinates}[1]{\textsc{in-coordinates}(#1)}
\newcommand{\leastcommon}{\textsc{leastcommon}}
\newcommand{\link}{\textsc{link}}
\newcommand{\cut}{\textsc{cut}}
\newcommand{\cycle}{\mathcal{C}}
\newcommand{\parent}{\textsc{parent}}
\newcommand{\cost}{\textsc{cost}}
\newcommand{\dcost}{\textsc{d-cost}}
\newcommand{\rcost}{\textsc{r-cost}}
\newcommand{\update}{\textsc{t-updatecost}}
\newcommand{\flip}[1]{\textsc{flip}(#1)}

\newcommand{\Bonichon}{colored }

\newcommand{\triangulation}{\mathtt{T}}
\newcommand{\flipgraph}{\mathcal{T}}
\newcommand{\fliprealizers}{\mathcal{R}}

\begin{document}
\title{Dynamic Schnyder Woods}
%\thanks{Supported by organization x.}}
\author{Sujoy Bhore\inst{1}\orcidID{0000-0003-0104-1659} \and
Prosenjit Bose \inst{2}\orcidID{0000-0002-8906-0573} \and
Pilar Cano\inst{3}\orcidID{0000-0002-4318-5282} \and
Jean Cardinal\inst{3}\orcidID{0000-0002-2312-0967} \and
John Iacono\inst{3,4}\orcidID{0000-0001-8885-8172}}
\authorrunning{S. Bhore, P. Bose, P. Cano, J. Cardinal, and J. Iacono.}
\institute{Indian Institute of Science Education and Research Bhopal, India.  \email{sujoy.bhore@gmail.com}\and Carleton University, Canada. \email{jit@scs.carleton.ca} \and
Université libre de Bruxelles, Belgium.
\email{\{pilar.cano, jean.cardinal\}@ulb.be}\\ \and
New York University, USA. \email{jiacono@ulb.be}}
\maketitle              % typeset the header of the contribution
\begin{abstract} 
A \emph{realizer}, commonly known as Schnyder woods, of a triangulation is a partition of its interior edges into three oriented rooted trees. A \emph{flip} in a realizer is a local operation that transforms one realizer into another. Two types of flips in a realizer have been introduced: \emph{\Bonichon flips} and \emph{cycle flips}. 
A corresponding \emph{flip graph} is defined for each of these two types of flips. The vertex sets are the realizers, and two realizers are adjacent if they can be transformed into each other by one flip.  In this paper we study the relation between these two types of flips and their corresponding flip graphs. 
%In particular, 
We show that a cycle flip can be obtained from linearly many \Bonichon flips. We also prove an upper bound of $O(n^2)$ on the diameter of the flip graph of realizers defined by \Bonichon flips. 
In addition, a data structure is given to dynamically maintain a realizer over a sequence of \Bonichon flips which supports queries, including getting a node's barycentric coordinates, in $O(\log n)$ time per flip or query. 


\end{abstract}
\section{Introduction}
Schnyder in his seminal work %developed an approach to the problem of constructing a plane straight line embedding of a planar graph by placing the vertices on a grid of small size%
proved that every planar graph with $n\ge 3$ vertices has a plane straight-line drawing in an $(n-2)\times (n-2)$ grid~\cite{schnyder1989planar, schnyder1990embedding}. %He proved that every planar graph with $n\ge 3$ vertices has a plane straight-line drawing in an $(n-2)\times (n-2)$ grid. 
This result was achieved in two parts: First, it was shown that every maximal planar graph admits a decomposition of its interior edges into three trees, called \emph{realizer}; Then, by using the realizer, a straight line embedding can be achieved. The Schnyder tree partitions, commonly referred to as \emph{Schnyder woods}, are an important concept in the area of graph drawing and order dimension theory; see~\cite{bonichon2007convex, felsner2005posets, felsner2004lattice, felsner2011order, felsner2014order}. Schnyder woods have been widely used to obtain combinatorial and algorithmic results for a wide range of problems from various domains, e.g., geometric spanners~\cite{bonichon2010plane}, optimal encoding
and uniform sampling of planar maps~\cite{poulalhon2006optimal},
compact data structures~\cite{aleardi2018array}, grid drawing~\cite{bonichon2007convex, gonccalves2014toroidal, schnyder1990embedding}, etc. Moreover, the connection between Schnyder woods and orthogonal surfaces have been explored over the years; see~\cite{bonichon2010connections, gonccalves2014toroidal, felsner2008schnyder, felsner2008orthogonal}. %cite-felsner'a papers...
Recently, Castelli Aleardi~\cite{aleardi2019balanced} considered \emph{balanced Schnyder woods}, in which the number of incoming edges of each color at each vertex is balanced, and provided linear time heuristics. 
Realizers are useful in designing algorithms for problems in graph drawing, graph encoding and others; see~\cite{chuang1998compact, kant1996drawing, de1990draw}. Brehm~\cite{brehm20003} and Mendez~\cite{ossona1994orientations} investigated the suitable operations that transform a realizer of a graph to another realizer of the same graph (see also~\cite{wagner1936bemerkungen, bonichon2002wagner}).

%flips.. 
A \emph{flip} in a realizer is a local operation that transforms one realizer into another. Two types of flips in a realizer have been introduced: \emph{\Bonichon flip} and \emph{cycle flip} (see Section~\ref{sec:flips}). 
%describe them in simple words... 
A corresponding \emph{flip graph} is defined for each of these two types of flips, the vertex sets of which are the realizers, and two realizers are adjacent if they can be transformed into each other by one flip.  %\\
%our contribution
%In this paper we study the relation between these two types of flips and their corresponding flip graphs. We also study the dynamic aspect of flips in realizer. We develop a data structure to dynamically maintain a realizer over a sequence of \Bonichon flips

\paragraph{\textbf{Our Contribution.}}
%\textbf{\emph{Our Contribution.}} 
We describe Schnyder woods and related constructions in Section~\ref{sec:sw}. In Section~\ref{sec:flips} we show that if an edge $e$ admits a diagonal flip in a triangulation $\triangulation$, then there exists a realizer $R$ of $\triangulation$ where the oriented edge $e$ admits a \Bonichon flip in $R$ (Section~\ref{subsec:flippable}). Later, we show that a cycle flip can be obtained from linearly many \Bonichon flips (Section~\ref{subsec:cycleflip}). Using these two results, we prove an upper bound of $O(n^2)$ on the diameter of the flip graph of realizers defined by \Bonichon flips (Section~\ref{subsec:upperbound}). 
Finally, in Section~\ref{sec:DynamicSec} we present a data structure to dynamically maintain a realizer under \Bonichon flips while supporting queries to a corresponding straight line embedding over a sequence of \Bonichon flips in $O(\log n)$ time per update or query. 
%while obtaining several static queries in $O(\log n)$ time. 

\section{Schnyder Woods}\label{sec:sw}

In this section we define a realizer and two other structures that are a bijection with realizers. 

%%%%%%%%%%%%%%%%%%%%%%%%%%%%%%%%%%%%%%%
\begin{figure}[t]
\centering
\includegraphics[page=13]{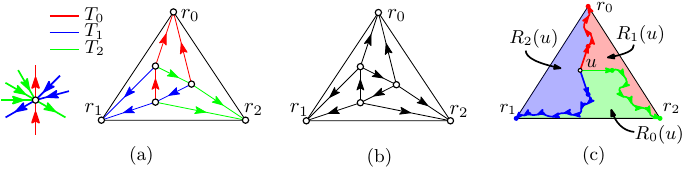}
\caption{(a) Example of the counter-clockwise order of the edges entering and leaving the vertex (\emph{left}) and a realizer (\emph{right}). (b) A 3-orientation. (c) Each colored region with its respectively colored path represents a region $R_i(u)$.}
\label{fig:contact-representation}
\end{figure}
%%%%%%%%%%%%%%%%%%%%%%%%%%%%%%%%%%%%%%%%

A \emph{triangulation} is a maximal planar graph (all faces are triangles), with a fixed outer face. 
\begin{definition}\label{def:realizer}
A \emph{realizer}, of a triangulation $\triangulation$ is a partition of its interior edges into three sets $T_0, T_1$ and $T_2$ of directed edges such that for each interior vertex $u$ the following holds:
\begin{compactenum}
\item Vertex $u$ has out-degree exactly one in each of $T_0, T_1$ and $T_2$ in counter-clockwise order.
\item All incoming edges of $T_i$ adjacent to $u$ occur between the outgoing edge of $T_j$ and $T_k$ for distinct $i,j,k \mod 3$. See Fig.~\ref{fig:contact-representation}(a).
\end{compactenum}
\end{definition}

Each tree $T_i$ of a realizer has as root $r_i$, one of the vertices in its outer face, and each vertex in the outer face is a sink in the directed graph defined by the given realizer. Note that a realizer can be represented as a $3$-coloring of its interior edges. See Fig.~\ref{fig:contact-representation}(a).
Schnyder defined realizers of triangulations in~\cite{schnyder1989planar,schnyder1990embedding} and proved that any triangulation with $n\geq 3$ vertices has a realizer. 

\begin{definition}
A \emph{3-orientation} of a triangulation $\triangulation=(V\cup\{r_0,r_1,r_2\}, E)$ is an orientation of the edges of $T$ such that each vertex has out-degree 3 except three special vertices $r_0,r_1,r_2$ that are sinks and define the outer face of $\triangulation$. See Fig.~\ref{fig:contact-representation}(b)
\end{definition}

In~\cite{de2001topological} de Fraysseix and de Mendez showed that any triangulation $\triangulation$ admits a $3$-orientation of its interior edges and that
the realizers of a triangulation $\triangulation$ form a bijection  with $3$-orientations of $\triangulation$.

\begin{definition}
A \emph{barycentric representation}\footnote{Note that this is called \emph{weak} barycentric representation in~\cite{schnyder1990embedding}.} of a triangulation $\triangulation$ is an injective function $u \in V(T)\mapsto (u_0,u_1,u_2) \in R^3$ that satisfies the conditions: 
\begin{compactenum}
\item $u_0+u_1+u_3=1$ for all vertices $u \in V(t)$. 
\item For each edge $uv$ and each vertex $w \notin \{u,v\}$, there is some $i \mod 3$ such that $(u_i, u_{i+1}) \prec (w_i, w_{i+1})$ and $(v_i, v_{i+1}) \prec (w_i, w_{i+1})$, where $\prec$ represents the lexicographic order.
\end{compactenum}
\end{definition}

For each interior vertex $u$ of $\triangulation$ we denote by $P_i(u)$ the path in $T_i$ from $u$ to its root $r_i$ with $i \mod 3$. For each interior vertex $u$ its paths $P_0(u), P_1(u)$ and $P_2(u)$ divide the triangulation into three disjoint regions $R_0(u), R_1(u)$ and $R_2(u)$ where $R_i(u)$ denotes the region defined by the vertices in path $P_{i+1}(u)\setminus\{u\}$ and the interior vertices enclosed by paths $P_{i-1}(u)$ and $P_{i+1}(u)$. See Fig.~\ref{fig:contact-representation}(c). The following lemma about these regions was shown in~\cite{schnyder1990embedding}.

\begin{lemma}[Schnyder~\cite{schnyder1990embedding}]\label{lemma:schnyder-regions}
For every different pair of interior vertices $u$ and $v$ of a triangulation it holds that if $v \in R_i(u)\cup P_{i-1}(u)$, then $R_i(v) \subset R_i(u)$.
\end{lemma}

Let $|R_i(u)|$ and $|P_i(u)|$ denote the number of vertices in $R_i(u)$ and $P_i(u)$, respectively. Let $f:V(\triangulation)\mapsto \mathbb{R}^3$ be the function defined as follows. For each interior vertex $u$ in $\triangulation$, $f(u)=\frac{1}{n-1}(|R_0(u)|, |R_1(u)|,$ $|R_2(u)|)$, and for each root $r_i \in T_i$, $f(r_i)$ has its $i$th coordinate equal to $n-1$, its $(i+1)$th coordinate equal to $1$ and its $(i+2)$th coordinate equal to $0$. Schnyder~\cite{schnyder1990embedding} showed that $f$ defines  barycentric coordinates of the vertices of $\triangulation$. Thus, every triangulation admits a barycentric representation that is in correspondence with a realizer.

\section{Flips}\label{sec:flips}

%%%%%%%%%%%%%%%%%%%%%%%%%%%%%%%%%%%%%%%%%%
\begin{figure}[h!]
\centering
\includegraphics[page=12]{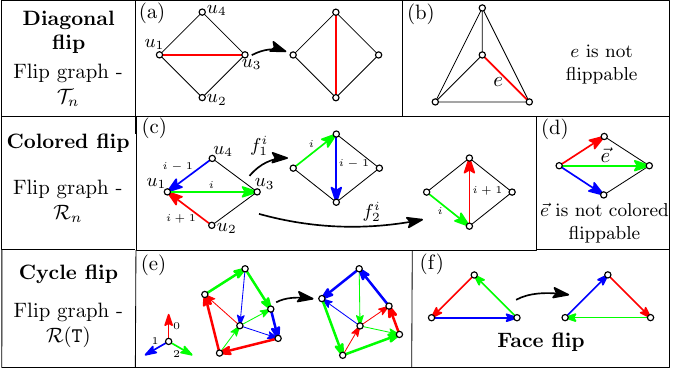}
\caption{Illustration of a diagonal flip, colored flip, cycle flip and face flip.}
\label{fig:flips}
\end{figure}
%%%%%%%%%%%%%%%%%%%%%%%%%%%%%%%%%%%%%%%%%%%

In this section we study the relationship between a \emph{diagonal flip} in triangulations with $n\geq4$ vertices,  a \emph{\Bonichon flip} in a realizer and a \emph{cycle flip} of a realizer of a triangulation. See Fig.~\ref{fig:flips}.
%The section is divided as follows. We first define  both type of flips and its variants. Then, we show that for any flippable edge in the triangulation there exists a realizer in which the edge is \Bonichon flippable and an algorithm for finding such realizer from an initial realizer in $O(n)$. Later on we describe how a cycle flip is related to \Bonichon flips and show that any cycle flip is a sequence of \Bonichon flips. Finally, we show an upper bound on the \Bonichon flip distance between two realizers.

%\subsection{Types of flips} 

A \emph{diagonal flip} in a triangulation $\triangulation$ is the operation that exchanges the diagonal $u_1u_3$ of a quadrilateral $u_1u_2u_3u_4$ in $\triangulation$ by the diagonal $u_2u_4$. See Fig.~\ref{fig:flips}(a). The \emph{flip graph of triangulations} ${\flipgraph}_n$ of $n$ vertices is defined as the graph with vertex set defined by all distinct triangulations on $n$ vertices and two vertices of $\flipgraph_n$ are adjacent if their corresponding triangulations can be transformed into each other by exactly one diagonal flip. We say that the diagonal $u_1u_3$ of a quadrilateral $u_1u_2u_3u_4$ in $\triangulation$ is \emph{flippable} if the edge $u_2u_4$ is not in $\triangulation$. See Fig.~\ref{fig:flips}(b).

Wagner~\cite{wagner1936bemerkungen} showed that the flip graph ${\flipgraph}_n$ is connected: Any triangulation of $n$ vertices can be transformed into another by a finite sequence of diagonal flips.
Given that a realizer is an orientation of the edges of a triangulation, it is natural to ask whether these flips can be extended to realizers. In other words, whether there exist local functions that transform one realizer into another. Bonichon et al.~\cite{bonichon2002wagner} defined flips in realizers that map to diagonal flips in the underlying triangulation.

We refer to edge $\Vec{uv}$ of a realizer of a triangulation $\triangulation$ as the oriented edge $uv$ of $\triangulation$.

\begin{definition}\label{def:bonichon-flip}
A \emph{\Bonichon flip} in a realizer of edge $\Vec{u_1u_3}$ with respect to $\Vec{u_2u_1}$ of the quadrilateral $u_1u_2u_3u_4$ is the operation that exchanges the edges $\Vec{u_1u_3}$ and $\Vec{u_2u_1}$ in tree $T_i$ and $T_j$, by the edges $\Vec{u_1u_2}$ and $\Vec{u_2u_4}$, respectively. 
There are two types of \Bonichon flips denoted $f_1^i$ or $f_2^i$ given in Fig.~\ref{fig:flips}(c). 

A \emph{\Bonichon flippable} edge preserves all realizers after it is flipped. 
\end{definition}

 Note that there might be edges that are flippable in $\triangulation$ that are not \Bonichon flippable as shown in Fig.~\ref{fig:flips}(d). However, Bonichon et al.~\cite{bonichon2002wagner} showed that the \Bonichon flip graph, denoted $\fliprealizers_n$, is connected. Their proof relies on the fact that the flip graph restricted to \Bonichon flips of the type $f_1^i$ defines a bounded poset as does the one restricted to their inverse $f_2^i$. %However, the proof does not implies a bound on the diameter of the graph.  

For simplicity, we only refer to a directed cycle as a \emph{cycle} in a directed graph. Brehm~\cite{brehm20003} defines cycle flips between $3$-orientations of a given triangulation, with its corresponding definition for realizers. 

\begin{definition}\label{def:cycle-flip}
A \emph{cycle flip} in a realizer $R$ is the operation that reverses the orientation of a cycle $\cycle$ such that if $\cycle$ is counter-clockwise oriented (resp. clockwise oriented), then: 
\begin{compactenum}
\item the color of each edge in $\cycle$ is exchanged by the color succeeding (resp. preceding) its original color,
\item for each edge inside $\cycle$ the new color is set to be the color preceding (resp. succeeding) its original color, and 
\item the color of all other edges is unchanged. See Fig.~\ref{fig:flips}(e). 
\end{compactenum}
A \emph{face flip} of a realizer $R$ is a cycle flip of a cycle of length $3$ defined by the edges of a face. See Fig.~\ref{fig:flips}(f). 
\end{definition}

Brehm~\cite{brehm20003} showed that given a 4-connected triangulation $\triangulation$, the flip graph of face flips $\fliprealizers(\triangulation)$ of the realizers of $\triangulation$ is connected. The proof is obtained by showing that the structure of flipping counter-clockwise faces into clockwise defines a poset, similar to the proof of \Bonichon flips. %In fact, he shows that the face flip is a lattice.

In this section, we provide a new proof of the connectivity of $\fliprealizers_n$ using the relation between \Bonichon flips and cycle flips. In addition, we prove an upper bound on the diameter of $\fliprealizers_n$. In order to show that $\fliprealizers_n$ is connected we divide this section as follows. In Subsection~\ref{subsec:flippable} we show that for any flippable edge $uv$ in a triangulation $\triangulation$ there exists a realizer $R$ of $\triangulation$ that admits a \Bonichon flip in edge $uv$. In Subsection~\ref{subsec:cycleflip} we show that any cycle flip in a realizer $R$ can be obtained by a sequence of a linear number of \Bonichon flips. In Subsection~\ref{subsec:upperbound}, we conclude, using the results from the two previous subsections, that two realizers $R$ and $R'$ can be transformed into each other by $O(n^2)$ \Bonichon flips. 

%%%%%%%%%%%%%%%%%%%%%%%%%%%%%%%%%%%%%%%%%%%%%%%%%
\subsection{Diagonal flips and \Bonichon flips}\label{subsec:flippable}

In this subsection we show that there exists a realizer $R$ of a triangulation $\triangulation$ for each flippable edge $e$ in $\triangulation$ in which the oriented edge $e$ in $R$ is \Bonichon flippable.

\begin{lemma}\label{lemma:find-cycle}
Let $\triangulation$ be a triangulation and $R$ be a realizer of $\triangulation$. Let $uv$ be a flippable edge in $\triangulation$ where $uv$ is the diagonal of the quadrilateral $uwvz$. If $\Vec{uv}$ is not \Bonichon flippable in $R$, then there exists a cycle $\cycle$ in $R$ that passes through either $\Vec{uw}$ or $\Vec{uz}$ in $R$ but avoids $\Vec{uv}$, that can be found in $O(n)$ time. 
\end{lemma}

\begin{proof}
Note that if either the edge $\Vec{wu}$ or $\Vec{zu}$ is in $R$, then $\Vec{uv}$ is \Bonichon flippable in $R$ and the result holds. Thus, both $\Vec{uw}$ and $\Vec{uz}$ are in $R$. See Fig.~\ref{fig:paths}(a). 
Let $i \mod 3$ be the label of $\Vec{uv}$ in $R$, hence the labels of $\Vec{uw}$ and $\Vec{uz}$ are $i+1$ and $i-1$ module $3$, respectively. 

%%%%%%%%%%%%%%%%%%%%%%%%%%%%%%%%%%%%%%%%%
\begin{figure}[t!]
\centering
\includegraphics[page=10]{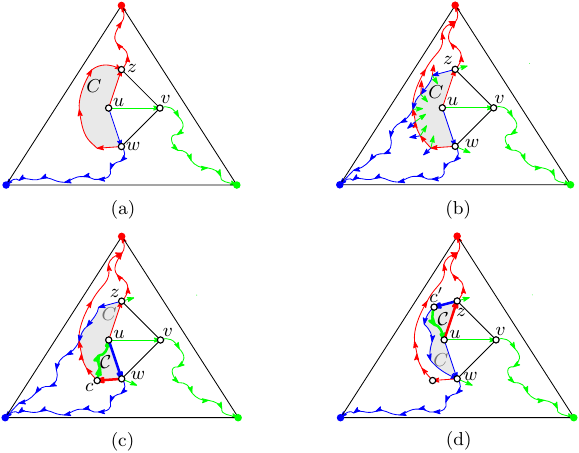}
\caption{An illustration of Theorem~\ref{thm:flippable}: (a) Vertex $z$ is in $P_{i+1}(w)$. (b) Vertex $z$ is not in $P_{i+1}(w)$. (c) The heavier cycle $\mathcal{C}$ when $(P_{i+1}(w)\setminus\{w\})\cap C \neq \emptyset$. (d) The heavier cycle $ \cycle $ when $(P_{i+1}(w)\setminus\{w\})\cap C = \emptyset$.}
\label{fig:paths}
\end{figure}
%%%%%%%%%%%%%%%%%%%%%%%%%%%%%%%%%%%%%%%%%%

Since $uv$ is flippable, it follows that at least one of $w$ or $z$ is an interior vertex of $\triangulation$. Assume without loss of generality that such vertex is $w$. Consider the paths $P_{i+1}(w)$ and $P_{i+1}(u)$. Note that both paths are of length at least 1, since both $u$ and $w$ are interior vertices of $\triangulation$. Since $w \in P_{i-1}(u)$, by Lemma~\ref{lemma:schnyder-regions} the path $P_{i+1}(w)$ lies in $R_i(u)$. For the same reason, if $z$ is an interior vertex of $\triangulation$, then the path $P_{i-1}(z)$ lies in $R_i(a)$ as well. See Figs.~\ref{fig:paths}(a) and~\ref{fig:paths}(b). Let $C$ be a closed region defined by the following boundary: 
If $P_{i+1}(w) \cap P_{i+1}(u) = z$, then $\partial C=w\cup P_{i+1}(w)\cup zuw$. See Fig.~\ref{fig:paths}(a). Otherwise, $\partial C= w \cup P_{i+1}(w)\cup(P_{i+1}(w)\cap P_{i-1}(z))\cup P_{i-1}(z)^{-1} \cup zuw$. See Fig.~\ref{fig:paths}(b). 

Note that by definition of realizer, each vertex in $\partial C \setminus\{u,w,z\}$ has its outgoing edge of $T_{i}$ in $C$. Similarly, by definition of realizer, each vertex in $(P_{i+1}(w)\cap\partial C)\setminus\{w\}$ and $( P_{i-1}(z)\cap\partial C)\setminus\{z\}$ has its outgoing edge of $T_{i-1}$ and $T_{i+1}$ outside $\mathcal{C}$, respectively. See Fig.~\ref{fig:paths}(b). 
In addition, by Definition of realizer, we have that an incoming edge from $T_i$ of an interior vertex $x$ in $\triangulation$ lies between its outgoing edges from $T_{i+1}$ and $T_{i-1}$. Thus, if a vertex in $\partial C\setminus\{u\}$ has an incoming edge from $T_{i}$, such edge is not in $C$. Therefore, for each vertex $x \in \partial C\setminus\{u,z,w\}$ its path $P_{i}(x)$ passes through $u$. 

Since $uv$ is flippable, it follows that $\partial C \setminus\{u,z,w\} \neq \emptyset$. Let $\cycle$ be a cycle in $R$ as follows: 
If $(P_{i+1}(w)\setminus\{w\}) \cap \partial C \neq \emptyset$, then let $\cycle= uwc\cup P_{i}(c)a$, which is a cycle in $R$ with $c$ the first vertex after $w$ in $P_{i+1}(w)$. See Fig.~\ref{fig:paths}(c). 
Otherwise, let $\cycle= uzc'\cup P_{i}(c')u$, which a cycle in $R$ with $c'$ the first vertex after $z$ in $P_{i-1}(z)$. See Fig.~\ref{fig:paths}(d). \qed
\end{proof}

Brehm~\cite{brehm20003} showed that applying a cycle flip to a realizer $R$ of a triangulation $\triangulation$ transforms $R$ into another realizer $R'$ of $\triangulation$. Thus, Lemma~\ref{lemma:find-cycle} implies the following.

\begin{theorem}\label{thm:flippable}
Let $e$ be a flippable edge in a triangulation $\triangulation$. Then, there exists a realizer $R$ of $\triangulation$ where the oriented edge $\Vec{e}$ in $R$ is \Bonichon flippable. 
\end{theorem}
\begin{proof}
Consider an arbitrary realizer $R$ of $\triangulation$ and let $\Vec{e}=\Vec{uv}$ be the orientation of $e$ in $R$. Let $uwvz$ be the quadrilateral in $\triangulation$ with diagonal $uv$ given in counter-clockwise order. If $\Vec{uv}$ admits a \Bonichon flip in $R$, then the statement holds. If $\Vec{uv}$ is not \Bonichon flippable, then from Lemma~\ref{lemma:find-cycle} there exists a cycle $\cycle$ that passes through either $\Vec{uw}$ or $\Vec{uz}$ in $R$ and does not pass through $\Vec{uv}$.

Let $R'$ be the oriented graph when flipping $C$. Hence, either $\Vec{wu}$ or $\Vec{zu}$ is in $R'$. In addition, edge $\Vec{uv}$ is in $R'$, since it does not lie in the interior of $C$ in $R$. By Brehm~\cite{brehm20003} the directed graph $R'$ is a realizer of $\triangulation$ different from $R$. Therefore, $\Vec{uv}$ is \Bonichon flippable in $R'$. \qed 
\end{proof}

%%%%%%%%%%%%%%%%%%%%%%%%%%%%%%%%%%%%%
\subsection{Cycle flips and \Bonichon flips}\label{subsec:cycleflip}

In this subsection we will show that any cycle flip of a realizer results from an $O(n)$ sequence of \Bonichon flips.

We say that a triangle $\triangle$ (not necessarily a face) in $\triangulation$ is \emph{three-colored} in $R$ if each pair of edges have different colors. From the Definition~\ref{def:realizer} we observe the following.

\begin{observation}\label{obs:three-colored}
A triangle $\triangle$ is three-colored in $R$ if and only if $\triangle$ is a cycle in $R$.
\end{observation}

Now, we show that a face flip results from two \Bonichon flips.

\begin{lemma}\label{lemma:2-consecutive-flips}
Let $F$ be an oriented face in a realizer $R$. Then, $F$ can be face flipped by two consecutive \Bonichon flips. 
\end{lemma}

\begin{figure}[tb]
\centering
\includegraphics[page=11]{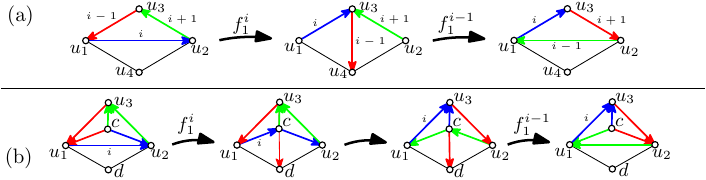}
\caption{(a) Illustration of Lemma~\ref{lemma:2-consecutive-flips}. (b) Illustration of Lemma~\ref{lemma:re-orienting-ST}.}
\label{fig:lemma-2-flips}
\end{figure}

\begin{proof}
Consider the oriented face $F=u_1u_2u_3$ in the realizer $R$. Let $R'$ be the realizer obtained when $F$ is face flipped by $F'=u_3u_2u_1$. Without loss of generality assume that $F$ is counter-clockwise oriented and that edge $u_1u_2$ has label $i \mod 3$ in $R$. 
Since $F$ is a three-colored triangle, all $u_1,u_2$ and $u_3$ are interior vertices. Hence, $u_1u_2$ is the diagonal of a quadrilateral $u_1u_4u_2u_3$, see Fig.~\ref{fig:lemma-2-flips}(a). Note that edge $u_1u_2$ is flippable in $\triangulation$: otherwise, either $u_2$ or $u_1$ is a vertex of degree $3$ in $\triangulation$ but with less than three outgoing edges. Which contradicts that $R$ is a realizer. 
Since $F$ is counter-clockwise oriented, $R$ admits a $f_{1}^i$ \Bonichon flip in the edge $u_1u_2$ with respect to $u_3u_1$. Let $R^{2}$ be the resulting realizer when applying such flip. Thus, the orientation of $\Vec{u_3u_1}$ changes to $\Vec{u_1u_3}$ and is re-labelled by $i$. In addition, the edge $\Vec{u_1u_2}$ by $\Vec{v_3v_4}$ and with label $i-1$. 
Now, note that $R^2$ admits an $f_1^{i-1}$ \Bonichon flip in the edge $\Vec{u_3u_4}$ with respect to $\Vec{v_2v_3}$. Consider the resulting realizer $R^{3}$ when applying such \Bonichon flipped in $R^2$. Then, the edge $\Vec{u_2u_3}$ changes its orientation to $\Vec{u_3u_2}$ and label by $i-1$. In addition, the edge $\Vec{u_3u_4}$ is exchanged by $\Vec{u_2u_1}$ with label $i+1$. Note that the face $u_1u_2u_3$ is now clockwise oriented in $R^{3}$ and the label of each edge of $F$ in $R$ is labelled by its successor in $R^{3}$. Even more, since none other edge has been changed, it follows that $F$ was face flipped in $R^3$. Therefore, $R^3=R'$. 

The reverse follows from the fact that $f_2$ \Bonichon flips are the inverse of $f_1$ \Bonichon flips. \qed
\end{proof}

%%%%%%%%%%%%%%%%%%%%%%%%%%%%%%%%%%%%%%%%%%
%%%%%%%%%%%%%%%%%%%%%%%%%%%%%%%%%%%%%%%%%%%

%Fig.~\ref{fig:lemma-2-flips}(a) illustrates Lemma~\ref{lemma:2-consecutive-flips}. 
The following lemma will be important for the next results.

\begin{lemma}\label{lemma:Brehm}[Brehm~\cite{brehm20003}]
Let $\triangulation$ be a triangulation and let $\cycle$ be a counter-clockwise cycle (resp. clockwise cycle) of length at least $4$ in a realizer $R$ of $\triangulation$ that contains no separating triangles. Then, $\cycle$ can be cycle flipped by a sequence of face flips of its interior faces where each face is flipped exactly once and it is oriented counter-clockwise (resp. clockwise) before it is flipped.
\end{lemma}

Using Lemmas~\ref{lemma:2-consecutive-flips} and~\ref{lemma:Brehm} we obtain the next result.

\begin{corollary} \label{corollary:2faces-flip}
Let $\triangulation$ be a triangulation and let $\cycle$ be a cycle of length $\geq 4$ in a realizer $R$ of $\triangulation$ with $m$ interior faces and it does not contain separating triangles. Then, $\cycle$ can be cycle flipped by a sequence of $2m$ \Bonichon flips.
\end{corollary}

Note that from Obs.~\ref{obs:three-colored} and definition of realizer, follows that the interior edges adjacent to the vertices of a cycle that is a separating triangle in $R$ are incoming edges. Using this fact we observe the following.

\begin{observation}\label{obs:flippable-ST}
Let $\cycle=u_1u_2u_3$ be a counter-clockwise (resp. clockwise) cycle  that is a separating triangle in a realizer $R$. Consider its interior face $cu_ju_{j+1}$ for any $j \mod 3$. Then, $R$ admits an $f_1^i$ (resp. $f_2^i$) \Bonichon flip in the edges $cu_j$ and $u_ju_{j+1}$.
\end{observation}

\begin{lemma}\label{lemma:re-orienting-ST}
Let $\cycle$ be a cycle that is a separating triangle in a realizer $R$ with $m$ interior faces and no interior separating triangles. Then, $\cycle$ can be cycle flipped by a sequence of $2m$ \Bonichon flips.
\end{lemma}
\begin{proof}
Let $\cycle = u_1u_2u_3$ be a counter-clockwise (resp. clockwise) cycle and consider its interior face $u_1cu_2$ and $\Vec{u_1u_2}$ with label $i \mod 3$. From Obs.~\ref{obs:flippable-ST} we can apply a \Bonichon flip to $u_1u_2$ with respect to $\Vec{cu_1}$ from a quadrilateral $u_1cu_2d$ in $\triangulation$. Let $R'$ be the resulted realizer when applying such \Bonichon flip. See Fig.~\ref{fig:lemma-2-flips}(b). Note that the edges $\Vec{u_1c}, \Vec{cu_2}, \Vec{u_2u_3}$ and $\Vec{u_3u_1}$ define a counter-clockwise cycle (resp. clockwise cycle) in $R'$ with $m-1$ interior faces. Let $\cycle '$ be such cycle. Let $R''$ be the resulted realizer when applying a cycle flip to $\cycle'$. From Corollary~\ref{corollary:2faces-flip} it  follows that $R''$ is obtained from $R'$ by a sequence of $2(m-1)$ \Bonichon flips.

Note that all interior edges of $\cycle'$ and edge $\vec{cu_1}$ are labelled in $R''$ by its preceding label (resp. succeeding label) in $R$. Also, edges $u_2u_3, u_3u_1$ and $cu_2$ have opposite orientation in $R''$ and its label in $R''$ is the succeeding (resp. preceding label) from $R$. Finally, $R''$ admits a \Bonichon flip in $\Vec{cd}$ with respect to edge $\Vec{u_2c}$. Let $R^3$ be the resulted realizer when applying such \Bonichon flip. Note that $\Vec{cu_2}$ had label $i$ in $R$ as $\Vec{u_1u_2}$ in $R$. See Fig.~\ref{fig:lemma-2-flips}(b). Hence, $R^3$ corresponds to a cycle flip of $\cycle$ in $R$. Therefore, $\cycle$ can be cycle flipped by a sequence of $2m-2+2=2m$ \Bonichon flips. \qed
\end{proof}

We say that a separating triangle in a cycle $\cycle$ is \emph{maximal} if it is not contained in another separating triangle contained in $\cycle$. Next, we show that any cycle flip is a sequence of $2m$ \Bonichon flips.

 %Using Lemmas~\ref{lemma:2-consecutive-flips},~\ref{lemma:Brehm} and~\ref{lemma:re-orienting-ST}, we prove the next lemma by induction on the number of separating triangles enclosed by a cycle $\cycle$.

\begin{lemma}\label{lemma:cycles}
Let $\cycle$ be a cycle with $m$ interior faces in a realizer $R$. Then, $\cycle$ can be cycle flipped by a sequence of $2m$ \Bonichon flips.
\end{lemma}
\begin{proof}
If $\cycle$ does not contain a separating triangles, then from Corollary~\ref{corollary:2faces-flip} and Lemma~\ref{lemma:re-orienting-ST} the statement holds.
Now, let us assume that $\cycle$ contains $t\geq 1$ separating triangles. 

Let us show that $\cycle$ can be cycle flipped by a sequence of $2m$ \Bonichon flips. 
We proceed by induction on $t$.

First, if $\cycle$ has length at least $4$, we denote $R' = R$ and $\cycle' =\cycle$ a cycle in $R'$ and $m'=m$. If $\cycle=u_1u_2u_3$ is of length $3$, i.e., is a separating triangle, we define $R'$, $\cycle'$ and $m'$ as follows. Consider its interior face $cu_1u_2$ and label $i$ of $\vec{u_1u_2}$. From Obs.~\ref{obs:flippable-ST} we can apply a \Bonichon flip to $u_1u_2$ from a quadrilateral $u_1cu_2d$ in $\triangulation$ such that $\vec{cu_1}$ is re-oriented to $\vec{u_1c}$ with label $i$ and $\vec{u_1u_2}$ is exchanged by $\vec{cd}$ with same label as $\vec{cu_1}$ in $R$. Let $R'$ be the resulted realizer when applying such \Bonichon flip to $R$. Note that the edges $\vec{u_1c}, \vec{cu_2}, \vec{u_2u_3}$ and $\vec{u_3u_1}$ define a cycle in $R'$ and is oriented as $\cycle$. Let $\cycle '$ be such cycle. Note that $\cycle'$ contains $m'=m-1$ interior faces and at most $t$ separating triangles.

[Base case] Assume $t=1$. Consider the interior separating triangle $\triangle$ of $\cycle'$ and let $m''$ be the number of interior faces in $\triangle$. Let $\cycle''$ be the resulting cycle when removing the interior vertices in $\triangle$. %Consider the resulting realizer $R''=R'\setminus{\cycle'}\cup\cycle$. 
From Lemma~\ref{lemma:Brehm}, $\cycle''$ can be cycle flipped from $R''$ by a sequence of $m' - m''+1$ many face flips.  Note that when applying the \Bonichon flips for face flipping each face $F$ in $\cycle''$, the rest of the edges are unchanged. In addition, since each face is face flipped exactly once on the same direction as $\cycle'$, it follows that replacing the face flip defined by $\triangle$ in $\cycle''$ by a cycle flip of $\triangle$, the interior edges in $\triangle$ have the corresponding new labelling.  Thus, $\cycle'$ can be cycle flipped by a sequence of $m'-m''$ face flips and a cycle flip of $\triangle$. From Lemmas~\ref{lemma:2-consecutive-flips} and~\ref{lemma:re-orienting-ST} we obtain that $\cycle'$ can be cycle by a sequence of $2m'$ \Bonichon flips.

If $\cycle$ is of length at least $4$, the statement holds. Otherwise, note since $\cycle'$ had the same orientation as $\cycle$, then when applying the cycle flip to $\cycle'$ we obtained $cu_2u_3u_1$ in the resulted realizer $R''$, where all of its interior edges are interior edges in $\cycle$ and were re-label as desired. Similarly, the orientation and labels of $u_1u_3$ and $u_2u_3$ were changed as desired. In addition, edge $cu_1$ was re-oriented twice (applying the \Bonichon flip to $u_1u_2$) with the corresponding labelling. It remains to re-orient edge $u_2c$ once and flip $cd$. In fact, $R''$ admits a \Bonichon flip in edges $\Vec{cd}$ and $\Vec{u_2c}$. Let $R^3$ the resulting realizer when applying the \Bonichon flip to $\Vec{cd}$. Note that $R^3$ contains the cycle $u_3u_2u_1$ and edge $\Vec{cu_2}$ with the desired labelling as for a cycle flip. Thus, $\cycle$ can be cycle flip by a sequence of $2+2m'=2m$ \Bonichon flips.

[Inductive Hypothesis] $\cycle$ can be cycle flipped by a sequence of $2m$ \Bonichon flips if it contains $t-1\geq 1$ separating triangles.

[Inductive step] Assume $\cycle'$ contains $t$ separating triangles. Let $\triangle_1, \ldots, \triangle_k$ be the maximal separating triangles in $\cycle'$ with $m_1, \ldots, m_k$ interior faces, respectively. Let $\cycle''$ be the resulting cycle when removing the interior vertices of each $\triangle_{1}, \ldots, \triangle_k$. Again, $\cycle''$ is a cycle of length at least $4$ with no separating triangles. By Lemma~\ref{lemma:Brehm}, $\cycle''$ can be cycle flipped by a sequence of $m'-(\sum_{j=1}^k m_j)+k$ face flips. Analogously as in the base case, we have that each face flip of a triangle $\triangle_j$ in $\cycle''$ corresponds as a cycle flip of $\triangle_j$ in $\cycle'$. Since each $\triangle_j$ contains at most $t-1$ separating triangles, by inductive hypothesis, we obtained that $\cycle'$ can be cycle flipped by a sequence of $2(m'-\sum_{j=1}^km_j)+2\sum_{j=1}^km_j = 2m'$ \Bonichon flips.

Analogously as in the base case, if $\cycle$ is of length at least $4$, the statement holds. Otherwise, since $\cycle'$ had the same orientation as $\cycle$, then when applying the cycle flip to $\cycle'$ we obtained $cu_2u_3u_1$ in the resulted realizer $R''$ and $R''$ admits a \Bonichon flip in edges $\Vec{cd}$ and $\Vec{u_2c}$. Let $R^3$ the resulting realizer when applying the \Bonichon flip to $\Vec{cd}$. Note that $R^3$ contains the cycle $u_3u_2u_1$ and edge $\Vec{cu_2}$ with the desired labelling as for a cycle flip. Thus, $\cycle$ can be cycle flip by a sequence of $2+2m'=2m$ \Bonichon flips. \qed 
\end{proof}

%%%%%%%%%%%%%%%%
\subsection{A bound on the diameter of $\fliprealizers_n$}\label{subsec:upperbound}

Komuro~\cite{komuro1997diagonal} proved that the diameter of $\flipgraph_n$ is $O(n)$ while the edges of the outer face are fixed~\footnote{The best bound known is in~\cite{cardinal2018arc} but their procedure might change the outerface.}. Using this and previous results we obtain the desire theorem.  

\begin{theorem}\label{thm:connectedness}
A given realizer of a triangulation of $n$ vertices is at \Bonichon flip distance $O(n^2)$ to any other realizer in $\fliprealizers_n$.
\end{theorem}

\begin{proof}
Let $R$ and $R'$ be two different realizers in $\fliprealizers_n$ and consider its underlying triangulations $\triangulation$ and $\triangulation'$, respectively. Let $\triangulation''$ be the triangulation with vertices $r_0$ and $r_1$ on its outerface adjacent to all the vertices. $\triangulation''$ has a unique realizer~\cite{brehm20003}. From Komuro~\cite{komuro1997diagonal}, $\triangulation$ and $\triangulation'$ can be transformed into $\triangulation''$ by $O(n)$ diagonal flips. Hence, from Theorem~\ref{thm:flippable} it follows that $R$ and $R'$ can be transformed into each other by a sequence of $O(n)$ \Bonichon flips and a cycle flip in between such flips. Since there can be cycles with $O(n)$ interior faces, from Lemma~\ref{lemma:cycles} it follows that $R$ and $R'$ can be transformed into each other by $O(n^2)$ \Bonichon flips. \qed
\end{proof}
%%%%%%%%%%%%%%%%%%%%%%
\section{Dynamic maintenance}\label{sec:DynamicSec}

In this section we study the problem of maintaining a realizer over a sequence of \Bonichon flips.

Let $T_i(u)$ denote the subtree of $T_i$ rooted at vertex $u$. Let $R$ be a realizer and let $R'$ be the resulted realizer when applying an $f_1^i$ (resp. $f_2^i$) \Bonichon flip to $\Vec{uv}$ with respect to edge $\Vec{wu}$ in quadrilateral $uwvz$. Define $c(u)=|R_{i-1}(w)|-|R_{i-1}(u)|+1$ (resp. $c(u)= |R_{i+1}(u)|-|R_{i+1}(w)|+1$) and define $c(w)$ as follows: If $\Vec{uz} \in R$, then $c(w)=-1$ (resp. $c(w)=0$). Otherwise, $c(w)=|R_{i}(u)|-|R_i(z)|$ (resp. $c(w)=|R_i(z)|-|R_i(u)|-1$).

Consider the labels $i$ and $j$ of $\Vec{uv}$ and $\Vec{wu}$, respectively. We observe that the only change made when applying a \Bonichon flip to $\Vec{uv}$ are the paths passing through edges $\Vec{uv}$ and $\Vec{uw}$. These paths are exactly the paths $P_i(x)$ for all $x \in T_i(u)$ and $P_{j}(y)$ for all $y \in T_{j}(w)$. Thus, the only vertices changing its regions are the ones in $T_i(u)$ and $T_{j}(w)$. Moreover, the $i$-th region of any $x \in T_i(u)$ and the $j$-th region of any $y \in T_j(w)$ remain unchanged. Thus, the regions $R_{i-1}(x)$ and $R_{i+1}(x)$ exchange elements for all $x \in T_i(u)$. The same applies to elements in $T_j(w)$. More precisely, we obtain the following lemma.
%for all $x \in T_i(u)$, we have that $|R'_{i-1}(x)|-|R_{i-1}(x)|=-(|R'_{i+1}(x)|-|R_{i+1}(x)|)=c(u)$ where $c(u)=|R'_{i-1}(u)|-|R_{i-1}(u)|$. Analogously, for all $y \in T_j(w)$, $|R'_{j-1}(y)|-|R_{j-1}(y)|=-(|R'_{j+1}(y)|-|R_{j+1}(y)|)=c(w)$ where $c(w)=|R'_{j-1}(w)|-|R_{j-1}(w)|$.
 
%%%%%%%%%%%%%%%%%%%%%%%%%%%%%%%%%%%%%%%%%%
\begin{figure}[tb]
\centering
\includegraphics[page=9]{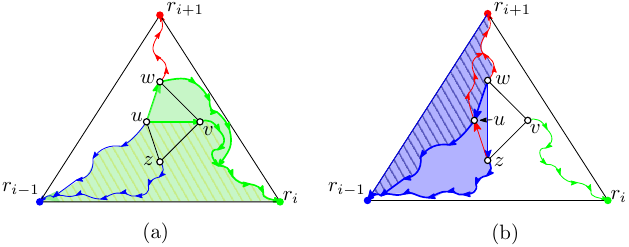}
\caption{(a) The filled area corresponds to $R'_{i+1}(u)=R_{i+1}(w)\cup\{w\}$ and the tilled one corresponds to $R_{i-1}(u)$. (b) The filled area corresponds to $R'_{i}(w)=R_i(w)\cup R_i(z)$ and the tilled area corresponds to $R_i(w)$.}
\label{fig:difference}
\end{figure}
%%%%%%%%%%%%%%%%%%%%%%%%%%%%%%%%%%%%%%%%%%%
%

\begin{lemma}\label{lemma:difference}
Let $R$ be a realizer and let $R'$ be the resulted realizer when applying a \Bonichon flip to $\Vec{uv}$ with respect to edge $\Vec{wu}$ in quadrilateral $uwvz$ with labels $i$ and $j$, respectively. Let $c(u)$ and $c(w)$ defined as above. Then, 
\begin{compactenum}
    \item For all $x$ in $T_i(u)$, $|R'_{i-1}(x)|-|R_{i-1}(x)|=c(u)$ and $|R'_i(x)|=|R_i(x)|$.
    \item For all $y$ in $T_{j}(w)$, $|R'_{j-1}(y)|-|R_{j-1}(y)|=c(w)$ and $|R'_{j}(y)|=|R_{j}(y)|$.
    \item The regions of any vertex in $V(R)\setminus (V(T_i(u)\cup T_{j}(w)))$ remain the same.
\end{compactenum}
\end{lemma}

\begin{proof}
Note that the only change made when applying a \Bonichon flip to $\Vec{uv}$ are the paths passing through edges $\Vec{uv}$ and $\Vec{uw}$. These paths are exactly the paths $P_i(x)$ for all $x \in T_i(u)$ and $P_{j}(y)$ for all $y \in T_{j}(w)$. 
Thus, the regions of any vertex in $V(R)\setminus (V(T_i(u)\cup T_{j}(w)))$ remain the same.

From Lemma~\ref{lemma:schnyder-regions} we note that each $x \in T_{i}(u)$ is in $R_i(u)$. Since the paths in $R_i(u)$ are unchanged in $R'$, it follows that $R'_i(x)=R_i(x)$ for all $x \in T_i(u)$. Similarly, we show that $R'_{j}(y)=R_{j}(y)$ for all $y \in T_{j}(w)$. 

Now, we assume $j=i-1$. Let us show that $|R'_{i-1}(x)|=|R_{i-1}(x)|+c(u)$. Since $u \in P_i(x)$ for each $x \in T_i(u)\setminus\{u\}$,  $u \in R_{i-1}(x)$. In addition, since $x \in R_i(u)$, $(R_{i-1}(x)\setminus R_{i-1}(u)) \subset R_i(u)$ which remains unchanged. Hence,  $|R'_{i-1}(x)|=|R_{i-1}(x)|-|R_{i-1}(u)|+|R'_{i-1}(u)|$. 
On the other hand, note that $P_{i-1}(u)\subset P_{i-1}(w)$. Hence, $R_{i+1}(u) \subset R_{i+1}(w)$. Note that $R'_{i+1}(u)$ is given by the region between paths $P_{i-1}(u)$ and $uw\cup P_i(w)$, which is exactly $R_{i+1}(w)\cup\{w\}$. See Fig~\ref{fig:difference}(a). Hence, $|R'_{i+1}(u)|-|R_{i+1}(u)|=|(R_{i+1}(w)\cup\{w\})\setminus R_{i+1}(u)| = - c(u)$. 
 Therefore, $|R'_{i-1}(x)|= |R_{i-1}(x)| + c(u)$ for all $x \in T_i(u)$. 

Finally, let us show that $|R'_{i+1}(y)|=|R_{i+1}(y)|+c(w)$ for all $y \in T_{i+1}(w)$. Since $w \in P_{i-1}(y)$ for each $y \in T_{i-1}(w)\setminus\{w\}$, $R_{i+1}(w) \in R_{i+1}(y)$. In addition, since $y \in R_{i-1}(w)$, $(R_{i+1}(y)\cap R_{i-1}(w)) \subset R_{i+1}(y)$ which remains unchanged. Hence,  $|R'_{i+1}(y)|=|R_{i+1}(y)|-|R'_{i+1}(w)|+|R_{i+1}(w)|$. 

On the other hand, if $\Vec{uz} \in R$: then $P_{i-1}(w) = (wvz)\cup P_{i-1}(z)$. Since $v$ is the only new vertex in the interior of $R'_{i+1}(w)$ and $|P'_{i+1}(w)|-|P_{i+1}(w)|=-1$, it follows that  $|R'_{i+1}(w)|-|R_{i+1}(w)|=-1$. 
Now, consider the case $\Vec{zu} \in R$: then $P_{i+1}(u) \subset P_{i+1}(z)$. Hence, $R_i(u) \subseteq R_i(z)$. In addition, since $\Vec{wu} \in R$, we have that $P_{i-1}(u)\subset P_{i-1}(w)$ and $R_i(u) \subseteq R_i(w)$. Thus, $R_i(z)\cap R_i(w)=R_i(u)$. Therefore,   $|R'_{i+1}(w)|-|R_{i+1}(w)|= - |R_i(z)\setminus R_i(w)|=|R_i(u)|-|R_i(z)|=c(w)$. See Fig~\ref{fig:difference}(b). Therefore, $|R'_{i+1}(y)|=|R_{i+1}(y)|+c(w)$ for every $y \in T_{i-1}(w)$.

The case when $j=i+1$ is symmetric. \qed
\end{proof}

A \emph{link/cut tree} is a data structure proposed by Sleator and Tarjan~\cite{sleator1983data} that maintains a forest of vertex-disjoint rooted trees with cost in its vertices under two dynamic operations: $\link$ and $\cut$ (see table below). %Each of these operations takes $O(\log n)$ time per update, where $n$ is the number of vertices. %Roughly speaking each tree is partitioned into vertex-disjoint paths, called \emph{solid} paths, and the remaining edges are called \emph{dashed}. Each solid path is represented by a splay tree and dashed edge to another tree. See Fig~\ref{fig:lct}. 
The link/cut tree supports the operations 1--6 from table below in worse case $O(\log n)$ time.

\subsubsection{The data structure.} Consider a data structure of a realizer $R$ as a set of three link/cut trees defined by each tree $T_0, T_1, T_3$. In each vertex $u$ we store its parent $\parent_i(u)$ in $T_i$ for each $i \mod 3$, its initial barycentric coordinates $\incoordinates{u}$ and two costs: $\dcost_i(u)$ and $\rcost_i(u)$. Where $\dcost_i$ refers to the distance of $u$ to the root $r_i$ of $T_i$ and the $\rcost_i$ refers to the difference between the initial ($i-1$)th coordinate with the current ($i-1$)th coordinate of $u$. Precisely, $\rcost_i(u)$ is the amount that has to be added to the initial ($i-1$)th coordinate and subtracted to the initial ($i+1$)th coordinate. The initial $\rcost$ is $0$. We define extra functions for our data structure in lines 7--11 from the table below. 
Using this data structure we obtain the next theorem. \\

\begin{tabularx}{0.96\textwidth} {|c |c |>{\raggedright\arraybackslash}X| }
 \hline
 1 & \link ($u, v$) & Add edge $uv$. \\
 \hline
 2 & \parent($u$)  & Return parent of vertex $u$.  \\
\hline
 3 & \cut($u$)  & Delete edge $v$\parent($v$).  \\
\hline
4 & \cost($u$)  & Return current cost in $u$  \\
\hline
5 & \update($u, x$)  & Add $x$ to the cost of all vertices in subtree $T(u)$.  \\
\hline
6 & \leastcommon($u,v$) & Return least common ancestor in $T$ of $u$ and $v$.\\
\hline
7 & \lab{$u,v$}  & Return label of edge $uv$.  \\
\hline
8 & \orientation{$u, v$}  & Return orientation of edge $uv$.  \\
\hline
9 & \incoordinates{$u$}  & Return initial barycentric coordinates of vertex $u$.  \\
\hline
%10 & $\dcost_i(u)$  & Return distance of $u$ to $r_i$ in tree $T_i$.  \\
%\hline
10 & \coordinates{$u$}  & Return current barycentric coordinates of $u$.  \\ 
\hline
11 & \flip{$u, v,w,z$}  & Apply \Bonichon flip to edge $\Vec{uv}$ with respect to $\vec{wu}$.  \\
\hline
\end{tabularx}

\

\begin{theorem}\label{thm:dynamic-realizer}
A realizer of a triangulation can be maintained in $O(\log n)$ per \textsc{flip}. Furthermore, queries \textsc{orientation}, \textsc{label}, \textsc{coordinates}, $\leastcommon_i$ and $\dcost_i$ can be obtained in $O(\log n)$ amortized time.
\end{theorem}
\begin{proof}
Consider the procedures define below. 

From~\cite{sleator1983data} $\parent$ and \textsc{in-coordinates} take $O(1)$ time and $\update_i$ takes worse case $O(\log n)$ time.
Since $|R_{i+1}(u)|=n-1-|R_i(u)|-|R_{i-1}(u)|$, it follows that $\coordinates{u}$ is correct. Since \textsc{in-coordinates} takes constant time and $\rcost$ was called exactly three times, it follows that \textsc{coordinates} can be obtained in $O(\log n)$ time.
Since the functions \textsc{orientation} and \textsc{label} are calling $\parent$ at most 6 times,  \textsc{orientation} and \textsc{label} can be obtained in $O(1)$ time. 
Since $\dcost_i$ and $\rcost_i$ behave as $\cost$ from a link/cut tree, then both can be obtained in $O(\log n)$. 
It remains to analyse \textsc{flip}. Note that in line 18 the function removes edges $\Vec{uv}$ and $\Vec{wu}$. In line 19 the new edges $\Vec{uw}$ and $\Vec{wz}$ are added. Thus, \textsc{flip} does the desired \Bonichon flip. Line 20 changes the $\rcost$ and $\dcost$ for each vertex in the subtree $T_i(u)$ by $c(u)$ and $d(u)$, respectively. Similarly, in the subtree $T_{j}(w)$ by $c(w)$ and $d(w)$, respectively. From Lemma~\ref{lemma:difference} updated $\rcost_i$ is correct. Therefore, \textsc{flip} is correct. Finally, we call exactly 3 times the function \textsc{coordinates}, twice each function $\cut, \link$ and $\update$. Each of these functions have amortized cost $O(\log n)$. Hence, \textsc{flip} has amortized cost $O(\log n)$. \qed 
\end{proof}

\noindent\fbox{\scalebox{0.75}{\begin{minipage}{\textwidth}
\begin{algorithmic}[1]
	 \Procedure{Label}{$u, v$}
	 \Comment{Returns the label of edge $uv$.}
	    \For{Each $i \mod 3$}
	        \If{$\parent_i(u) = v$}
	        \State return $i$
	        \Else 
	        \If{$\parent_i(v) = u$}
	        \State \Return $i$
	        \EndIf
	    \EndIf
	    \EndFor
	    \EndProcedure
\end{algorithmic}
\end{minipage}}}

\noindent\fbox{\scalebox{0.75}{\begin{minipage}{\textwidth}
\begin{algorithmic}[1]
	 \Procedure{orientation}{$u, v$}
	 \Comment{Returns orientation of edge $uv$.}
	    \State Let $b= \textsc{False}$
	    \For{each $i \mod 3$}
	        \If{$\parent_i(u) = v$}
	        \State let $b= \textsc{True}$
	    \EndIf
	    \EndFor
	    \If{$b= \textsc{True}$}
	    \State \Return $\Vec{uv}$
	    \Else
	    \State \Return $\Vec{vu}$
	    \EndIf
	    \EndProcedure
\end{algorithmic}
\end{minipage}}
}

\noindent\fbox{\scalebox{0.75}{\begin{minipage}{\textwidth}
\begin{algorithmic}[1]
	 \Procedure{Coordinates}{$u$}
	 \Comment{Returns a vector with the barycentric coordinates of $u$ in current realizer $R$.}
		\State let $(u_0, u_1, u_3)=\incoordinates{u}$.
		\For{each $i \mod 3$}
		    \State let $c_i = \cost_i{u}$.
		 \EndFor
		 \For{each $j \mod 3$}
		    $u'_j = u_j + \frac{c_{j+1}-c_{j-1}}{n-1}$.
		 \EndFor
		 \State \Return $(u'_0, u'_1, u'_2)$
		 \EndProcedure
\end{algorithmic}
\end{minipage}}
}

\noindent\fbox{\scalebox{0.8}{\begin{minipage}{\textwidth}
\begin{algorithmic}[1]
	 \Procedure{Flip}{$u, v, w, z$}
	 	\Comment{Creates a flip while updates the cost in the subtrees that are changed.}
			\State Let $i=\lab{u,v}, j=\lab{u,w}$.
			\State let $d(u)=\dcost_i(w)-\dcost_i(u)+1$ and $d(w)=\dcost_j(z)-\dcost_j(w)+1$
			\State $(v_0, v_1, v_2)= \coordinates{u}$, 
			\State $(w_0, w_1, w_2)=\coordinates{w}$, 
			\State $(z_0, z_1, z_2)=\coordinates{z}$.
			\If{$j=i-1 \mod 3$}
			    \State let $c(u) = (n-1)(w_{i-1}-u_{i-1})$.
				\If{$\orientation{u,z} =\Vec{uz}$}
				\State let $c(w)= 0$
				\Else \State let $c(w)=(n-1)(z_i-u_i)$
				\EndIf
			    \Else \State let $c(u) = (n-1)(u_{i-1}-w_{i-1})+1$
			   \If{$\orientation{u,z}=\Vec{uz}$}
			        \State let $c(w)=-1$
			        \Else \State let $c(w)=(n-1)(u_i-z_i)$
			       \EndIf
			\EndIf
			\State $\cut_i(u), \cut_j(w)$
			\State $\link_i(u, w), \link_j(w, z)$
			\State $\update_i(u, c(u), d(u)), \update_j(w, c(w), d(w))$.
	\EndProcedure
\end{algorithmic}
\end{minipage}}
}
 \bibliographystyle{splncs04}
 \bibliography{Arxiv.bbl}

\begin{thebibliography}{10}
\providecommand{\url}[1]{\texttt{#1}}
\providecommand{\urlprefix}{URL }
\providecommand{\doi}[1]{https://doi.org/#1}

\bibitem{bonichon2007convex}
Bonichon, N., Felsner, S., Mosbah, M.: Convex drawings of 3-connected plane
  graphs. Algorithmica  \textbf{47}(4),  399--420 (2007)

\bibitem{bonichon2010connections}
Bonichon, N., Gavoille, C., Hanusse, N., Ilcinkas, D.: Connections between
  theta-graphs, {D}elaunay triangulations, and orthogonal surfaces. In:
  International Workshop on Graph-Theoretic Concepts in Computer Science. pp.
  266--278. Springer (2010)

\bibitem{bonichon2010plane}
Bonichon, N., Gavoille, C., Hanusse, N., Perkovi{\'c}, L.: Plane spanners of
  maximum degree six. In: International Colloquium on Automata, Languages, and
  Programming. pp. 19--30. Springer (2010)

\bibitem{bonichon2002wagner}
Bonichon, N., Le~Sa{\"e}c, B., Mosbah, M.: Wagner's theorem on realizers. In:
  International Colloquium on Automata, Languages, and Programming. pp.
  1043--1053. Springer (2002)

\bibitem{brehm20003}
Brehm, E.: 3-orientations and schnyder 3-tree-decompositions. Master's thesis,
  FB Mathematik und Informatik, Freie Universit{\"a}t Berlin  (2000)

\bibitem{cardinal2018arc}
Cardinal, J., Hoffmann, M., Kusters, V., T{\'o}th, C.D., Wettstein, M.: Arc
  diagrams, flip distances, and hamiltonian triangulations. Computational
  Geometry  \textbf{68},  206--225 (2018)

\bibitem{aleardi2019balanced}
Castelli~Aleardi, L.: Balanced schnyder woods for planar triangulations: an
  experimental study with applications to graph drawing and graph separators.
  In: International Symposium on Graph Drawing and Network Visualization. pp.
  114--121. Springer (2019)

\bibitem{aleardi2018array}
Castelli~Aleardi, L., Devillers, O.: Array-based compact data structures for
  triangulations: Practical solutions with theoretical guarantees. Journal of
  Computational Geometry  \textbf{9}(1),  247--289 (2018)

\bibitem{chuang1998compact}
Chuang, R.C.N., Garg, A., He, X., Kao, M.Y., Lu, H.I.: Compact encodings of
  planar graphs via canonical orderings and multiple parentheses. In:
  International Colloquium on Automata, Languages, and Programming. pp.
  118--129. Springer (1998)

\bibitem{de2001topological}
De~Fraysseix, H., de~Mendez, P.O.: On topological aspects of orientations.
  Discrete Mathematics  \textbf{229}(1-3),  57--72 (2001)

\bibitem{de1990draw}
De~Fraysseix, H., Pach, J., Pollack, R.: How to draw a planar graph on a grid.
  Combinatorica  \textbf{10}(1),  41--51 (1990)

\bibitem{felsner2004lattice}
Felsner, S.: Lattice structures from planar graphs. the electronic journal of
  combinatorics pp. R15--R15 (2004)

\bibitem{felsner2014order}
Felsner, S.: The order dimension of planar maps revisited. SIAM Journal on
  Discrete Mathematics  \textbf{28}(3),  1093--1101 (2014)

\bibitem{felsner2008orthogonal}
Felsner, S., Kappes, S.: Orthogonal surfaces and their cp-orders. Order
  \textbf{25}(1),  19--47 (2008)

\bibitem{felsner2011order}
Felsner, S., Nilsson, J.: On the order dimension of outerplanar maps. Order
  \textbf{28}(3),  415--435 (2011)

\bibitem{felsner2005posets}
Felsner, S., Trotter, W.T.: Posets and planar graphs. Journal of Graph Theory
  \textbf{49}(4),  273--284 (2005)

\bibitem{felsner2008schnyder}
Felsner, S., Zickfeld, F.: Schnyder woods and orthogonal surfaces. Discrete \&
  Computational Geometry  \textbf{40}(1),  103--126 (2008)

\bibitem{gonccalves2014toroidal}
Gon{\c{c}}alves, D., L{\'e}v{\^e}que, B.: Toroidal maps: Schnyder woods,
  orthogonal surfaces and straight-line representations. Discrete \&
  Computational Geometry  \textbf{51}(1),  67--131 (2014)

\bibitem{kant1996drawing}
Kant, G.: Drawing planar graphs using the canonical ordering. Algorithmica
  \textbf{16}(1),  4--32 (1996)

\bibitem{komuro1997diagonal}
Komuro, H.: The diagonal flips of triangulations on the sphere. Yokohama
  mathematical journal  \textbf{44}(2),  115--122 (1997)

\bibitem{ossona1994orientations}
Ossona~de Mendez, P.: Orientations bipolaires. Ph.D. thesis, Paris, EHESS
  (1994)

\bibitem{poulalhon2006optimal}
Poulalhon, D., Schaeffer, G.: Optimal coding and sampling of triangulations.
  Algorithmica  \textbf{46}(3),  505--527 (2006)

\bibitem{schnyder1989planar}
Schnyder, W.: Planar graphs and poset dimension. Order  \textbf{5}(4),
  323--343 (1989)

\bibitem{schnyder1990embedding}
Schnyder, W.: Embedding planar graphs on the grid. In: Proceedings of the first
  annual ACM-SIAM symposium on Discrete algorithms. pp. 138--148 (1990)

\bibitem{sleator1983data}
Sleator, D.D., Tarjan, R.E.: A data structure for dynamic trees. Journal of
  computer and system sciences  \textbf{26}(3),  362--391 (1983)

\bibitem{wagner1936bemerkungen}
Wagner, K.: Bemerkungen zum vierfarbenproblem. Jahresbericht der Deutschen
  Mathematiker-Vereinigung  \textbf{46},  26--32 (1936)

\end{thebibliography}
\newpage
\appendix

\end{document}